# Unveiling the origin of unconventional moiré ferroelectricity

Ruirui Niu[1#], Zhuoxian Li[1#], Xiangyan Han[1#], Qianling Liu[1], Zhuangzhuang Qu[1], Zhiyu Wang[1], Chunrui Han[2,3*], Kenji Watanabe[4], Takashi Taniguchi[4], Kaihui Liu[1], Jinhai Mao[5], Wu Shi[6,7], Bo Peng[8], Zheng Vitto Han[9,10*], Zizhao Gan[1], Jianming Lu[1*]

[1]State Key Laboratory for Mesoscopic Physics, School of Physics, Peking University, Beijing 100871, China

[2]Institute of Microelectronics, Chinese Academy of Sciences, Beijing, 100029, China

[3]University of Chinese Academy of Sciences, Beijing 100049, China

[4]National Institute for Materials Science, 1-1 Namiki, Tsukuba, 305-0044, Japan.

[5]School of Physical Sciences and CAS Center for Excellence in Topological Quantum Computation, University of Chinese Academy of Sciences, Beijing, China

[6]State Key Laboratory of Surface Physics and Institute for Nanoelectronic Devices and Quantum Computing, Fudan University, Shanghai 200433, China.

[7]Zhangjiang Fudan International Innovation Center, Fudan University, Shanghai 201210, China.

[8]National Engineering Research Center of Electromagnetic Radiation Control Materials, School of Electronic Science and Engineering, University of Electronic Science and Technology of China, Chengdu, China

[9]State Key Laboratory of Quantum Optics and Quantum Optics Devices, Institute of Opto-Electronics, Shanxi University, Taiyuan 030006, China

[10]Collaborative Innovation Center of Extreme Optics, Shanxi University, Taiyuan 030006, China

* Corresponding author: hanchunrui@ime.ac.cn, vitto.han@gmail.com, jmlu@pku.edu.cn

**Abstract**

**Interfacial ferroelectricity emerges in heterostructures consisting of non-polar van der Waals (vdW) layers, greatly expanding the scope of two dimensional ferroelectrics. In particular, the unconventional moiré ferroelectricity observed in bilayer graphene/boron nitride (BN) heterostructures, exhibits promising functionalities with topological current, superconductivity and synaptic responses. However, the debate about its mechanism – correlation driven charge transfer between two graphene layers – limits device reproducibility and hence large-scale production. Here by designing a single-layer graphene encapsulated by lattice-mismatched $WSe_2$, we identify the ferroelectricity as stemming from – instead of graphene moiré bands – the particular BN, where interfacial sliding ferroelectricity must play a role. With similar structures, multilayer twisted $MoS_2$ is found to reproduce the ferroelectricity. The key is a conductive moiré ferroelectric, where the screened gate and the pinned domain wall together result in unchanged electronic states, i.e. anomalous screening. The intimate connection to interfacial sliding ferroelectricity thus provides advantages of diverse choices of constituent materials and robust polarization switching while preserving the unique anomalous screening, paving the way to reproducible and reliable memory-based devices in artificial intelligence.**

**Introduction**

Artificial stacking of non-polar van der Waals materials into interfacial ferroelectrics opens a new paradigm of multi-functional vdW devices[1–6], allowing non-volatile switching of electrical/magnetic/optical properties of the non-polar constituent in a designed manner. Two branches of interfacial ferroelectricity have been developed: one is the lattice-driven sliding ferroelectricity that relies on non-centrosymmetric stacking of two layers[7–15], in which the polarization switching through interlayer sliding guarantees robust repeatability and durability; the other is the electron-driven moiré ferroelectricity observed in crystallographically aligned bilayer graphene/BN heterostructure[3,16–20], which was ascribed to the electronic correlation induced interlayer charge transfer[16,18]. Especially, the latter was recently shown to non-volatile control unconventional superconductivity and various exotic correlated phases[3,5]. Furthermore, it works as a synaptic transistor[19] that can perform diverse biorealistic neuromorphic functionalities, whose non-stochastic nature guarantees high repeatability and long-term endurance. However, the mechanism of unconventional moiré ferroelectricity remains debated[16–20]; the concomitant observation of anomalous screening that is believed to be the key is yet to be understood. The absent guidance of device construction leads to an ultralow yield (typically five out of one hundred devices in practice). To realize these promising applications which need a large array of devices, it is highly demanding to clarify the underlying mechanism.

The recent finding of coexisting charge carriers and polarization in vdW ferroelectrics such as few-layer $WTe_2$[7,21,22], $MoTe_2$[4] and multilayer twisted $MoS_2$[14] may provide a novel solution: (1) While the ferroelectric polarization can be switched by the gate, substantial charge carriers of the ferroelectric electrically screen the bottom gate, thus isolating the top detector. (2) The polarization field of ferroelectrics, as an effective intermediate gate, solely controls the detector's electronic states. All these factors, together with microscopic characterization of electric-field driven polarization dynamics[15,23,24], open new opportunities to tackle the long-standing puzzle in unconventional moiré ferroelectricity.

Here we report the unusual ferroelectricity observed in a BN/WSe$_2$/monolayer graphene/WSe$_2$/BN heterostructure with arbitrary twist angles, where the monolayer graphene (MLG) provides a simple spatial distribution of electron wavefunction and the WSe$_2$ encapsulation eliminates the formation of moiré superlattice. Both designs can safely exclude the electron-driven mechanisms such as correlation induced interlayer charge transfer and electron/hole ratchet that depend on the bilayer structure and correlated moiré bands. Consequently, the ferroelectricity must be from the rare and unusual BN flake, which can only be ascribed to interfacial sliding ferroelectricity. To substantiate the proposal, we construct a heterostructure consisting of monolayer graphene as a detector and multilayer twisted transition metal chalcogenides (MoS$_2$) as a part of the gate dielectric layer. The nearly identical behavior unveils the key ingredient – conductive moiré ferroelectrics, in which polar domains are arranged in an antiferroelectric pattern. In this scenario, the anomalous screening is owing to both the Coulomb screening of the gate and pinning of domain walls; the seemingly gate-working regions correspond to the electric field-driven polarization switching process instead of conventional electrostatic gating. Overall, our results merge the two branches of interfacial ferroelectricity: the unconventional moiré ferroelectricity originates from sliding ferroelectricity with finite conductance and hence inherits the advantages of non-damage polarization switching and fruitful material choices, providing a promising avenue to advanced functionalities that require memory-based devices in high density and large-size arrays.

## Results

### Ferroelectricity in monolayer graphene/BN heterostructures

While all previous studies focused on spatially distinguishable two-layer graphene, including Bernal[16–19], twisted[3] and BN intercalated bilayer[20] graphene, here we construct a heterostructure consisting of a monolayer graphene encapsulated by WSe$_2$ (Part A) or BN (Part B) (Fig. 1a). This is because (1) Monolayer graphene is a single sheet of two-dimensional electron gas with a well-defined spatial distribution along the out-of-plane direction. (2) WSe$_2$ has a large lattice mismatch (~30%) with graphene and BN, so any long-period moiré superlattices can be eliminated on the graphene or BN surface.

Fig. 1b-c shows distinct dependences on the top ($V_t$) and bottom ($V_b$) gates, where the hysteretic behavior is only observed for $V_t$ (See more monolayer devices in Fig. S4, S5). Firstly, unlike the bilayer graphene, electrons in monolayer graphene cannot be divided into two groups with different physical characteristics or layer distribution, which contradicts previous models such as correlation induced interlayer charge transfer[16] and excitonic ferroelectricity[18,19]. Secondly, almost identical results are found for Part A and B. On the one hand, it indicates WSe$_2$ plays no role to form ferroelectricity. On the other hand, due to the absence of moiré superlattices between graphene and WSe$_2$, it directly proves that a moiré superlattice

(at least, the long-periodic one that results in a narrow moiré band) is not necessary, which agrees with previous experiments[17] where the polarized charge density is well beyond the capacity of a moiré band. Actually, in most of ferroelectric graphene/BN heterostructures[3,16–20], one cannot see the superlattice gap manifested as a satellite resistance peak. In addition, the persistence to room temperature of the hysteresis is unlikely to be a correlation effect (See temperature dependence in Fig. S3, S4). All of these facts lead to a conclusion that the above ferroelectricity does not stem from the interface between graphene and BN. Instead, it originates from the BN itself.

More specifically, the ferroelectricity originates from the particular BN flake of the hysteretic gate. For clarity, we define the ferroelectric BN as Fe-BN to distinguish it from the normal hexagonal BN (h-BN). To support this argument, we plot a two dimensional colored map of resistances versus both gates in Fig. 1d. With abundant Landau levels, the anomalous screening region (the nearly vertical ridge in parallel with the $V_t$ axis; see strictly vertical ridges in Fig. S4, S5) can be identified for all electronic states as well as charge neutral points (CNP). Obviously, all these features depend on the magnitude of $V_t$ instead of a displacement field ($D$) involving contributions from both gates. We noticed that ferroelectric devices made of Bernal bilayer or rhombohedral hexalayer (SI, section 6) graphene have a similar dependence, which, again, is in contrast to the electron-driven mechanisms, because the layer polarization of electrons in these multilayer graphene depends on $D$ rather than a single gate.

In Fig. 1d, the ferroelectric gate $V_t$ is taken as the slow scanning axis, where the forward and backward scanning have no hysteresis (Fig. S1). Upon changing $V_t$ as a fast scanning gate, a hysteresis loop can be formed by subtracting the backward from the forward scanning (Fig. 1e). The hysteresis magnitude, i.e. the range of anomalous screening, is found to be significantly larger than that in Fig. 1d. Phenomenologically, the contrast is due to the relatively short relaxation time, i.e., the magnitude of charge polarization initiated by a large gate voltage decays with time by itself. As such, the non-hysteretic phase stays in equilibrated states, and the hysteresis loop must take into account dynamics of polarization switching (see Fig. S2, S4h and detailed explanation in Fig. 3).

**Reproduced ferroelectric behavior in multilayer twisted MoS$_2$**

So far, unconventional ferroelectric devices investigated have been mainly made of bulk Fe-BN flakes, which are rare among exfoliated flakes from high-quality crystals. By excluding charged defects as the cause of resistance hysteresis (SI, section 8), the only mechanism left for a ferroelectric BN is sliding ferroelectricity. The candidates of Fe-BN structure involve rhombohedral stacking and screw dislocation induced moiré superlattice. As the charge polarization of a rhombohedral BN is difficult to switch[25] (also see SI, section 7), the likely scenario is the screw dislocation[26–38], which was recently found to generate three dimensional moiré superlattices in spiral graphite and TMDs[26–28]. Unexpectedly, the mixed stacking order in BN has been reported to be conductive at gain boundaries[39].

Here we construct such ferroelectrics by stacking MoS$_2$ – structurally resembling hBN and easily obtained – with small twist angles for multiple times (Fig. 2a, see another device in Fig. S8). At each moiré interface, there is a network of triangular domains separated by topologically protected domain walls. Each domain is positively or negatively polarized along the out-of-plane direction due to the non-centrosymmetric configuration, which are illustrated in top and cross-section views. They are arranged in an antiferroelectric way, so that without external electric field the averaged polarization of these two polar domains are nulled. When the $E$ field exceeds a threshold value[23,24], e.g. $E_c^+$, the positively polarized domain starts to grow in area via domain wall motion, and the negatively polarized domain shrinks correspondingly.

Now we extend the above picture to multilayer MoS$_2$. Unlike the bilayer case, the uncontrolled variation of twisted angles between many interfaces makes it difficult to obtain exactly the same moiré pattern within the whole heterostructure. By omitting the details, we define the cumulative polarization as a single parameter $P$ (Fig. 2b). As MoS$_2$ is naturally electron doped by sulfur vacancies, the multilayer stack can be viewed as an additional conductive channel in parallel with graphene. On the one hand, when it is grounded by the graphene, the bottom gating field is expected to be screened and thus the graphene detector merely subjects to the polarization field of ferroelectric MoS$_2$, i.e., $P$, $-P$ or 0. On the other hand, the contribution of MoS$_2$ to the measured conductance can be omitted, owing to the much larger mobility and higher work function of graphene. In other words, the bottom gate switches ferroelectric polarization of twisted MoS$_2$, which sequentially controls graphene conductance.

To verify the proposed scenario, we constructed a heterostructure with hexalayer twisted MoS$_2$. As shown in Fig. 2c, a significant hysteresis could be observed. The reference sample (monolayer graphene on the same BN flake but without MoS$_2$, Fig. S6) does not show any hysteresis. The comparison indicates that the ferroelectricity indeed stems from the twisted MoS$_2$. The salient feature – anomalous screening – is presented in Fig. 2d. In the phase diagram, schematic polar domain networks are overlaid to explain the main features. (1) Within the range $[E_c^-, E_c^+]$ of $V_b$, the polar domains are equal along positive and negative directions, leading to a zero polarization field. As the gating field of $V_b$ is screened, the graphene detector is fully controlled by $V_t$, and always stays at CNP when $V_t$=0 (Fig. 2b). (2) Upon the gating field of $V_b$ goes beyond the threshold, domain walls will be driven to expand one of the polar domains, which results in nonzero averaged polarization that can be evidenced by the shift of CNP along the $V_t$ axis. We must emphasize that, in this regime the trace of CNP (or the slope of the high-resistance ridge) does not reflect the capacitance ratio between top and bottom gates. Rather, it is completely controlled by the domain wall motion.

**Charge polarization dynamics mediated by domain wall motion**

An important observation of the anomalous screening in Fig. 2d is that, the forward and backward scanning of $V_b$ do not exhibit a hysteresis loop (similar to that in Fig. S1d-f), which apparently contradicts with the strong hysteresis in Fig. 2c. The distinct behaviors are due to the extremely slow scanning rate in the former and much faster scanning in the latter, indicating that field driven domain wall is not simply determined by the magnitude of electric field. To fully understand the relation, we now examine the dynamics of domain wall.

The longitudinal resistance ($R_{xx}$) and Hall density ($n_H$) are recorded at $B$=1 T in Fig. 3a-e and Fig. 3f-j, respectively. In each loop, we first set a fixed negative charge polarization ($V_b$ swept to -10 V) and then scan $V_b$ to different positive values ($V_b^{max}$). Representative results are shown as follows (more data in Fig. S7): (1) For a small $V_b^{max}$=2 V, the charge polarization is always strongly negative within the full loop, leading to small resistances (Fig. 3a) and substantial hole carriers (Fig. 3f). This can be understood, since the positive field is not sufficiently large to trigger the domain wall motion for flipping the negative polarization. (2) With a larger $V_b^{max}$ =4.5 V (Fig. 3c), CNP cannot be accessed during the forward scanning (red curve), as also seen from the positive $n_H$ in the forward scanning (Fig. 3h). When $V_b$ is swept backwards (blue curve), the resistance continues to increase and quickly passes the CNP peak; correspondingly, the carrier type changes from hole to electron. A straightforward scenario is that, the domain wall starts to move under the large electric field and it does not stop even when $V_b$ turns back after reaching $V_b^{max}$. Interestingly, it is during the backward scanning that the domain wall passes the neutral position (denoted by the CNP peak and $n_H$=0). (3) Further increasing $V_b^{max}$>5 V, CNP can be easily reached in the forward scanning, as evidenced by the peak resistance (Fig. 3d, e) and zero Hall density (Fig. 3i, j).

We emphasize again that, the anomalous screening corresponds to the pinning of domain wall and the seemingly 'gate-working-region' corresponds to the depinning process. So in Fig. 3k, we can label the schematic domain patterns for a typical full hysteresis loop. Several features should be stressed: (1) Take the backward scanning as an example, the pinning of domain wall takes place whenever the scanning is reversed from forward to backward direction, which reflects the retention of the polarized domain. (2) Once the depinning process is initiated, the domain wall motion directly reverses the polarization without a wide zero-polarization region (i.e., $[E_c^-, E_c^+]$ in Fig. 2d), resulting a sharp CNP peak. This may be attributed to the inertia of domain wall. (3) For $V_b$<-6 V, the domain wall is not pinned, but moves with the magnitude of $V_b$, as seen from the varying $n_H$. However, it is nearly pinned in the window of [-4 V, 4 V]. Understanding of these details about pinning-depinning needs more investigation of domain wall dynamics[15,23,24,40].

**Ladders in the phase diagram**

An interesting feature of multilayer interfacial ferroelectricity is the quantized ladders of charge polarization, thanks to the strongly confined dipoles at each interface[14]. The critical electric field that switches the polar domains may vary at different interfaces. This is reasonable, since the twisted angle determines the dynamics of domain wall but it is difficult to keep the same angle in a multilayer moiré superlattices. Consequently, in our multilayer twisted $MoS_2$ a multilevel ladder is expected. In order to observe the full phase diagram, the gate scanning range is greatly expanded (Fig. 4a). Several features of anomalous screening can be found, indicating the pinning of domain walls in these regions. We thus propose a model of interface-specific switching of polar domains, which is schematically shown at every transition point. After polarizing all the interfaces, the graphene ends up with anomalous screening.

Nevertheless, the ladder can be suppressed in a uniformly stacked three dimensional moiré superlattice. Here we plot a characteristic phase diagram for ferroelectric BN heterostructures in Fig. 4b. Although the state is a little unstable under a very large electric field (top left and bottom right corners), one can identify that the graphene eventually enters into the region of anomalous screening. The lack of intermediate ladders suggests that the twisted angles are uniform across all relevant interfaces. Recently, three dimensional moiré superlattices have been directly synthesized taking advantages of screw dislocation[26–38]. With finely controlled twisted angle, such a spiral structure is promising to serve as a universal ferroelectric substrate with anomalous screening.

## Conclusion

In this work, by integrating monolayer graphene in heterostructures we firmly establish that the unconventional moiré ferroelectricity originates from some particular boron nitride flakes. With reproduced ferroelectricity in multilayer twisted $MoS_2$, interfacial ferroelectrics with finite conductance is identified as the key to explain all experimental features: While the gate is screened by this conductive layer, the graphene detector can only be tuned by the polarization field of the ferroelectric; the anomalous screening region is ascribed to static states of polarization (e.g. the pinning of domain walls), whereas the seemingly normal region actually reflects the switching process of polarization (e.g. domain wall movement).

The unification of interfacial ferroelectricity and ferroelectricity with anomalous screening is intriguing. The latter naturally inherits robust repeatability and durability of the former, while demonstrating unique properties that lead to promising applications in neuromorphic computing. Moreover, based on the knowledge of interfacial ferroelectricity, one can expand the constituent layer to a large family of non-polar van der Waals materials including the most studied boron nitride and abundant TMDs like $MX_2$ (M: Mo, W; X: S, Se, Te), as long as they are doped to be conductive. With recent advances in wafer-scale monolayer synthesis[41–43] and stacking techniques[28,44–46], ferroelectric devices with anomalous screening are ready to develop into a large-scale array for massive memory-based networks, laying the foundation of realizing real-word applications such as biomimicking functionalities and in-memory computing.

## Methods
**Sample fabrication**

Monolayer graphene, few-layer graphite gates, transition metal dichalcogenides and boron nitride flakes were all mechanically exfoliated from bulk crystals on to silicon wafers and identified with optical microscopy. Rhombohedral boron nitride flakes were grown following ref.[47]. Multilayer heterostructures were then fabricated following the standard dry transfer method. Then reactive ion etching was used to pattern the Hall bar geometry. At last, one-dimensional edge contacts were prepared following the standard e-beam lithography and e-beam evaporation of Cr (1 nm)/Au (50 nm).

**Electrical measurement**

All the transport measurements were carried out in a cryostat equipped with a superconducting magnet up to 9 T. A standard four-probe method of constant current was performed. The AC current was supplied by Stanford Research Systems SR830 lock-in amplifiers with a working frequency at 17.777 Hz. The DC gate voltages were output by two Keithley 2400 Source Meters.

## Data availability

All figures are provided in Source Data file. All other data that support the findings of this study are available from the corresponding author upon reasonable request.

## Acknowledgements

This work was supported by NSF of China (grant no. 11974027, 62275265), National Key R&D Program of China (grant no. 2019YFA0307800, 2021YFA1400100), Beijing Natural Science Foundation (grant no. Z190011, 4222084). We also acknowledge the support from Peking Nanofab.

## Contributions

J.L., Z.H. and C.H. conceived the project. R.N., Z.L. and X.H. fabricated devices and performed transport measurements with assistance from Z.Q. and Z.Y. Crystallographic characterization was performed by R.N., Z.L., Q.L. and T.L.; K.W. and T.T. synthesized boron nitride crystals. J.L., Z.H., C.H., J.M., K.L., Z.G. supervised the project. All authors contribute to the data analysis. R.N., Z.H., C.H. and J.L. wrote the paper with input from all authors.

## Competing interests

The authors declare no competing interests.

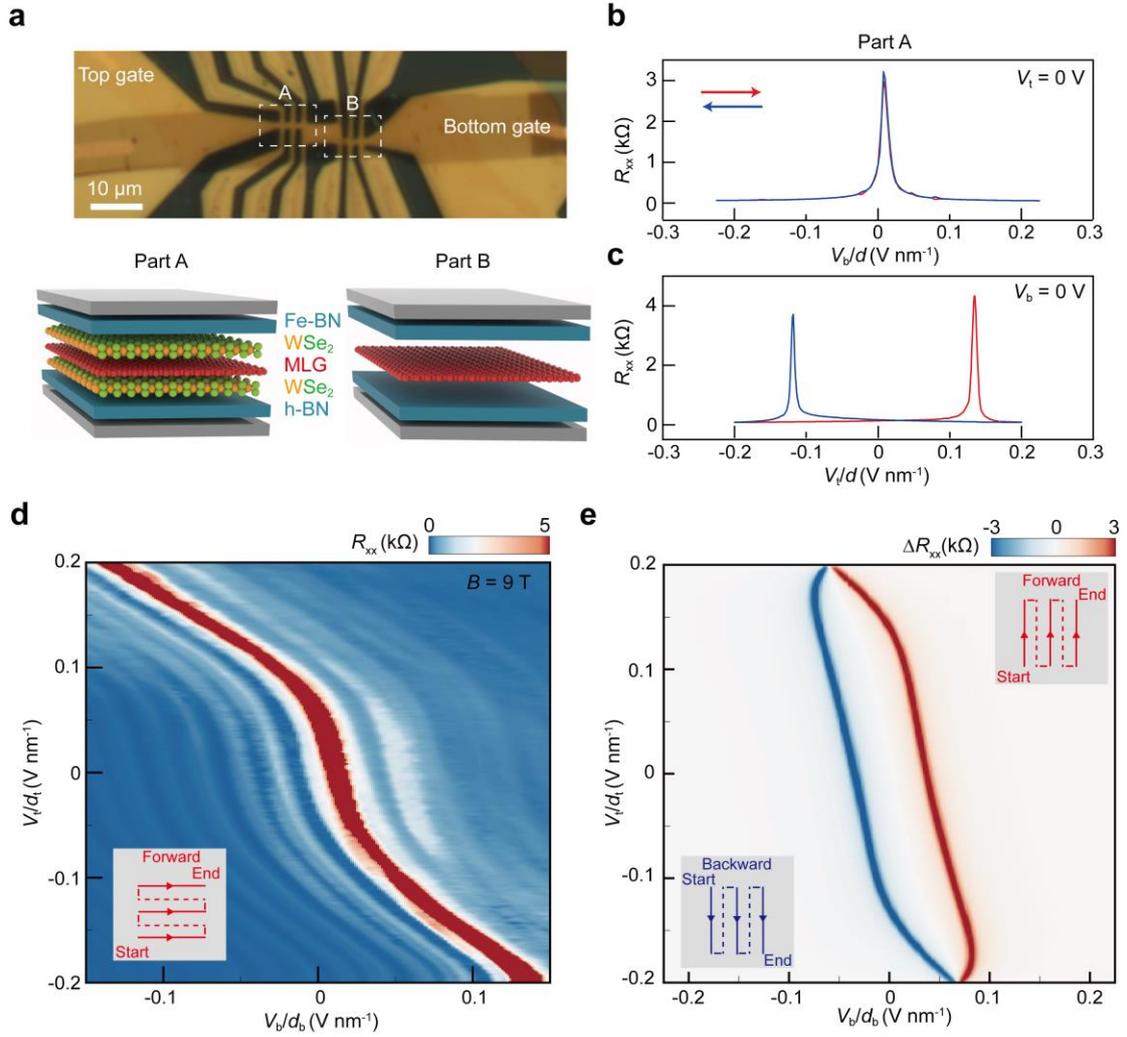

**Figure 1 Ferroelectricity in a monolayer graphene heterostructure. a.** Optical characterization of the device with two different parts with structure schematics. Note that the top BN that exhibits ferroelectricity has a different notation with the bottom BN. **b-c** For both Part A and B, hysteretic resistance curves are only observed by sweeping the top gate. **d.** Two dimensional resistance mapping with $B$=9 T (See the scanning configuration in the inset). The nearly vertical feature of the CNP trajectory represents the anomalous screening regime where $V_t$ seemingly stops to work. The Landau level spectroscopy also enables visualization of electronic states with low resistances. Obviously, the anomalous screening depends on the specific gate, i.e., $V_t$ in this device, rather than the displacement field that is contributed by both top and bottom gates. **e.** Hysteresis loop between forward and backward sweeping of the top gate. Insets: Two scanning configurations.

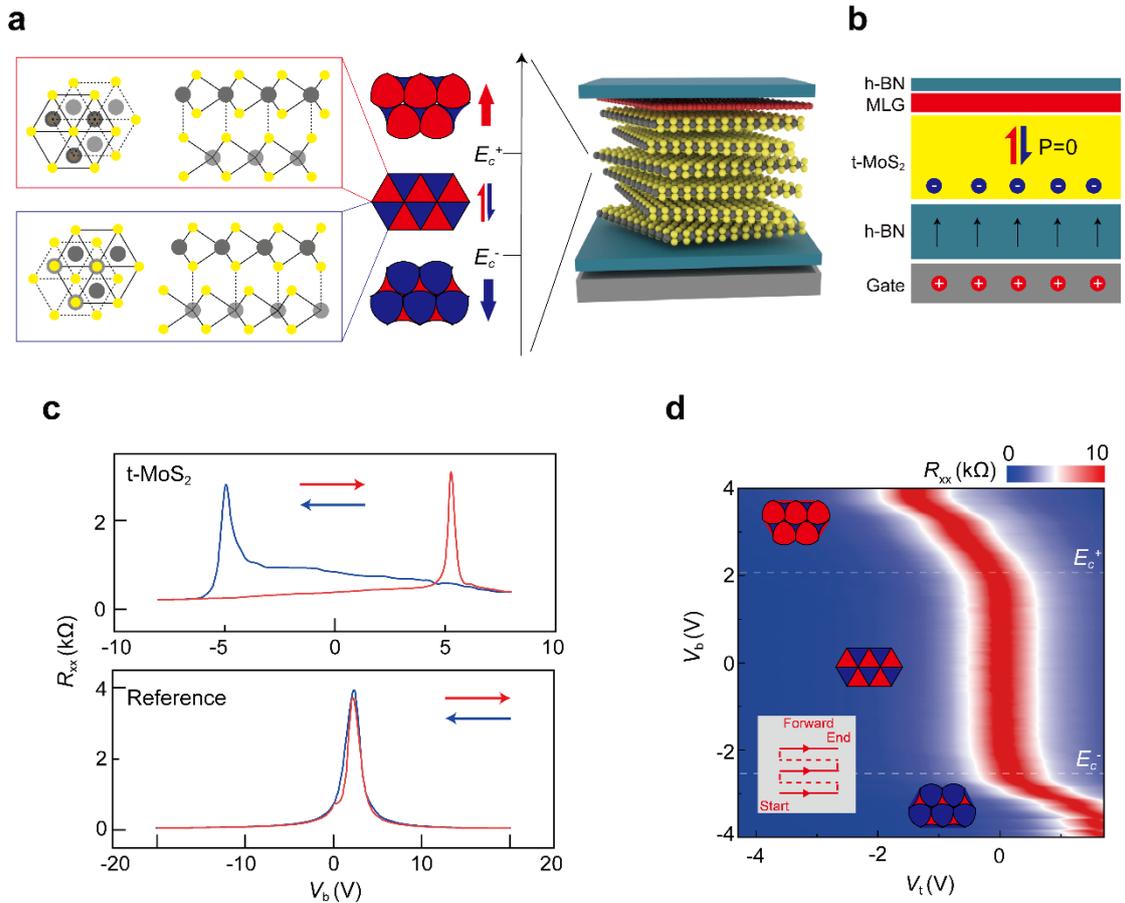

**Figure 2 Ferroelectricity observed in multilayer twisted MoS$_2$. a.** Device schematic of a graphene-MoS$_2$ heterostructure. For the twisted MoS$_2$, a typical moiré superlattice with antiferroelectric domains is depicted, for which atomic structures of the two polarized domains and their evaluation as a function of electric field are illustrated. Note that the critical fields $E_c$ denote the transition from antiferroelectricity to polarized states. **b.** Working mechanism of anomalous screening. The gating field is screened by the intrinsically electron-doped MoS$_2$, but can polarize the interfacial ferroelectrics made of MoS$_2$. The ferroelectric polarization field subsequently changes the state of the monolayer graphene on top. Here we show the case of $E_c^- < E < E_c^+$, where the graphene always stays at CNP. **c.** Top: Hysteretic transfer curves are observed when scanning back and forth $V_b$. We emphasize that the hysteresis is not resulted from a ferroelectric BN, since another part of graphene without twisted MoS$_2$ does not show any hysteresis (bottom). **d.** A full phase diagram showing the anomalous screening regime. The schematic polar domains for transitions are overlaid. Inset: the gate scanning scheme.

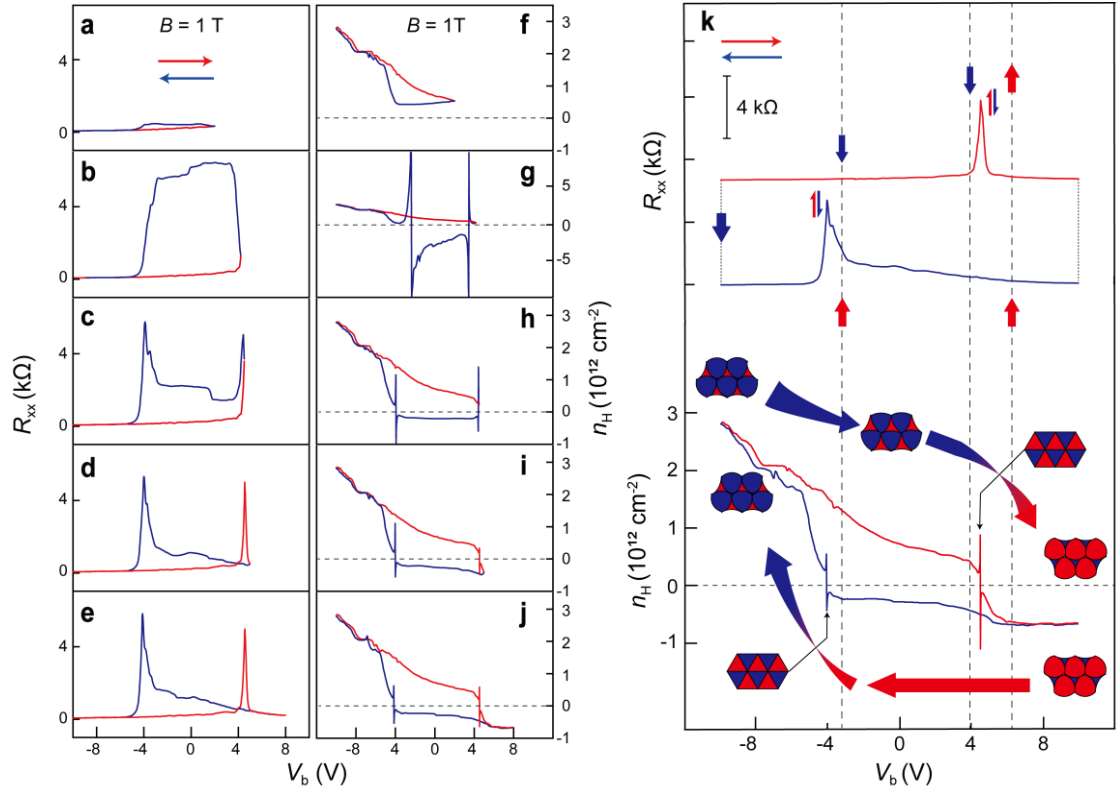

**Figure 3 Dynamics of polarization switching. a-j**. Longitudinal resistance (**a-e**) and Hall carrier density (**f-j**) characterization by varying the sweeping range of the positive gate. Here the negative limit of $V_b$ is fixed at -10 V and the positive limit is set as 2, 4, 4.5, 5, 8 V, respectively. While in **a** positive polarization is not achieved, the anomalous screening just happens on the CNP position in **b**. The domain wall dynamics can be best presented in **c**, where the CNP is not accessed during the forward scanning, but the electronic state continues to pass the CNP from hole doping to electron doping in the process of backward scanning (as evidenced in **h**). **k**. Illustration of varying charge polarization by arrows (top, $R_{xx}$) and polar domains (bottom, $n_H$) during a full sweeping range.

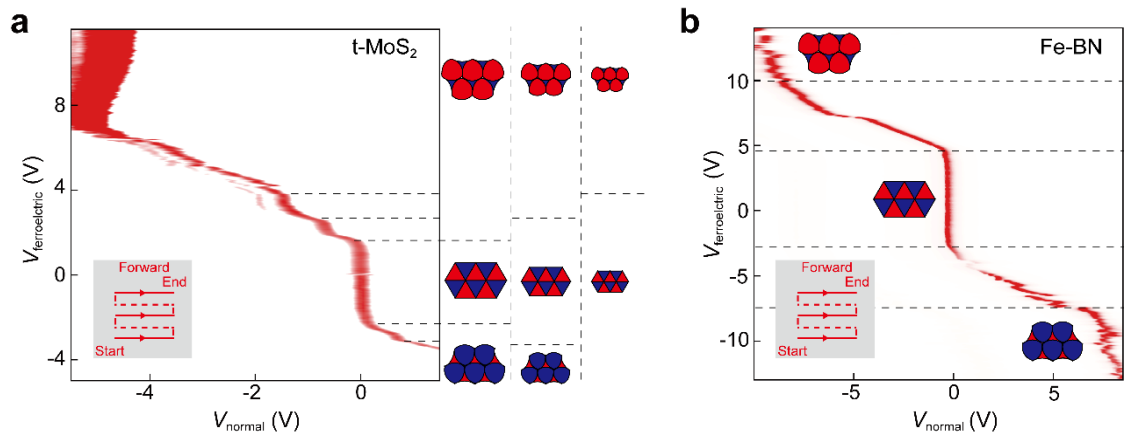

**Figure 4 Ladders of ferroelectricity. a.** Multiple ladders of anomalous screening observed in the phase diagram of the multilayer twisted $MoS_2$ device. Schematic is attached for layer-resolved switching of polar domain networks, where smaller domains require higher switching fields. **b**. Phase diagram for a typical BN-graphene heterostructure, in which fewer ladders indicate more homogenous distribution of polar domains.